\documentstyle[preprint,aps]{revtex}
\tighten 
\begin{document}
\draft
\preprint{\vbox{To be published in Phys. Lett. B 
\hfill  \hfill RCNP-Th00047}}

\title{Cubic Casimir operator of SU$_C$(3) and confinement 
in the nonrelativistic quark model}
\author{V. Dmitra\v sinovi\' c}         
\address{
Vin\v ca Institute of Nuclear Sciences \\
P.O.Box 522, 11001 Beograd, Yugoslavia; \\
address until July 2001: Research Center for Nuclear Physics,\\
Osaka University, Mihogaoka 10 - 1, Ibaraki, Osaka 567-0047, Japan }

\maketitle                 

\begin{abstract}
Only two-body [${\rm F}_{i} \cdot {\rm F}_{j}$] confining potentials 
have been 
considered, thus far, in the quark model without gluons, which by 
construction can only depend on the quadratic Casimir operator of the 
colour SU(3) group.  
A three-quark potential that depends on the cubic Casimir operator is 
added 
to the quark model.
This results in improved properties of $q^3$ colour non-singlet 
states, 
which can now be arranged to have (arbitrarily) higher energy than 
the singlet, and the ``colour dissolution/anticonfinement" problem of 
the ${\rm F}_{i} \cdot {\rm F}_{j}$ model is avoided. 
\end{abstract}
\widetext
  
\section{Introduction}  

In spite of the wide-spread concensus on the validity of QCD as 
the theory of strong interactions, 
QCD has proven essentially intractable, except in perturbative 
approximations. 
\footnote{The highest achievement of lattice QCD is the (mere) 
extraction
of a linearly rising colour-singlet $q \bar q$ potential, which is  
far from proving physical confinement.
In particular, the existence and energetics of coloured states remain 
unexplored. 
Moreover, multiple colour-singlet multi-quark states appear 
in the theory and it is unclear just which one lies lowest.}
It is fair to say that little (or no) 
understanding of quark confinement has been 
achieved since QCD's inception more than 25 years ago
\footnote{In this regard, see M. Chanowitz's remarks 
made 23 years ago \cite{Chanowitz}.} 

Instead of solving QCD one often resorts to various forms of the 
quark model, 
perhaps the simplest version being the nonrelativistic (n.r.) 
constituent quark model 
\footnote{One may think of it as QCD with gluon degrees of freedom 
frozen, the quark-quark interaction being transmitted by potentials of 
colour-exchange character.}.
The spin-statistics problem of the simplest quark model 
led to the introduction of the colour degrees of freedom that obey 
the SU(3) Lie algebra (this led subsequently to QCD). This 
``colour SU(3)" is exactly conserved 
(there is no colour leakage), hence the quark model spectrum must 
fall into
irreducible representations of this group, and the
quark Hamiltonian must be expressible in terms of SU(3) invariant 
operators.
There are two such independent ``invariants" of SU(3), the so-called
Casimir operators, other than the unit operator \cite{BB63,Pais66}.

Even with an infinitely rising (``confining") $q-q$ potential, that
does not allow separation of individual quarks from their aggregates, 
in the simplest quark model, which assumes colour independent 
quark interactions, the coloured $q^3$ states are degenerate with
the colour-singlet one. 
This is in manifest contradiction with the experience. To remedy 
this shortcoming, a 
colour-dependent factor ${\rm F}_{i} \cdot {\rm F}_{j}$ 
(proportional to the colour charges of the two quarks 
${\rm F}_{i,j}^{a}$) was introduced into the two-quark potential of 
the quark model, in analogy with the one-gluon exchange (OGE) potential 
in QCD. This goes by the name of the ``${\rm F}_{i} \cdot {\rm F}_{j}$ 
model". 
This colour 
factor is proportional to the first (``quadratic") Casimir operator 
$C^{(1)} = {\rm F}^{a} {\rm F}^{a} = {\rm F} \cdot {\rm F} = {\rm 
F}^2$, 
where ${\rm F}^{a}$ are the group generators of SU(3). 
No three-quark or higher-order interactions have been allowed in this 
model so far. Whereas many view three- and many-body forces with 
distaste, it is also indisputable that  
QCD demands three-, and four-quark interactions 
at the tree approximation level. Therefore it ought to have been clear 
all along that no two-body interaction could describe QCD completely.

\subsection{Problems with the ${\rm F}_{i} \cdot {\rm F}_{j}$ model} 

As noted by many, e.g. \cite{LeY90,Isgur}, the ${\rm F}_{i} \cdot {\rm 
F}_{j}$ model suffers from a number of weaknesses: 
\begin{enumerate}
\item {it predicts unstable coloured $q^2$ and $q {\bar q}$ states 
(the ``colour dissolution / anticonfinement" problem);} 

\item {it predicts towers of new, as yet undiscovered multiquark 
states, $q^2 {\bar q}^2$ being the lowest lying ones (the ``colour 
chemistry" problem);}

\item {it predicts unobserved long-range forces between colour 
singlets (the v.d. Waals force problem).} 
\end{enumerate}

The standard ``solution" to problem (1), the assumption 
that only colour singlet states exist, is entirely {\it ad hoc} and
thus unsatisfactory. Moreover, it does not 
begin to address problems (2) and (3). Hence we shall seek a change 
in the dynamics of the quark model 
that might lead to the solution of the confinement problem(s). 
We consider the displacement of coloured states
to (arbitrarily) high energies/masses as a solution to the 
confinement problem.

This note will show that the introduction of a three-quark force 
proportional to the second (``cubic") Casimir operator can fix (at 
least some of) these shortcomings. 
\footnote{The second Casimir operator $C^{(2)}$ of SU(3) is tri-linear 
(``cubic") in the group generators ${\rm F}^{a}$, {\it viz.} $C^{(2)} 
= d^{abc}{\rm F}^{a} {\rm F}^{b} {\rm F}^{c}$, so it can only appear 
in three- or more-quark potentials.} 
To be sure, such a three-quark 
potential is not an arbitrary addition: it arises from the
instanton-induced 't Hooft interaction in QCD. What we do assume, 
however, is that its spatial behaviour is confining, which is {\it ad 
hoc}. 

\section{Three-body potential} 
 
The three-quark potential can be factored into 
a colour part ${\cal C}_{123}$ and the spin-spatial part ${\cal V}_{123}$:
\begin{eqnarray}
V_{123} &=& {\cal C}_{123} {\cal V}_{123}
\label{e:3bpot} .\
\end{eqnarray}
The following 3-body colour factors can be written down:
\begin{eqnarray}
{\cal C}_{123} { } = \cases{
\sum_{i < j}^{3} {\rm F}_{i} \cdot {\rm F}_{j} = 
{\rm F}_{1} \cdot {\rm F}_{2} + {\rm F}_{1} \cdot {\rm F}_{3} +
{\rm F}_{2} \cdot {\rm F}_{3}
\label{e:sum} \cr
d^{abc}{\rm F}_{1}^{a} {\rm F}_{2}^{b} {\rm F}_{3}^{c}  
\label{e:d}\cr
i f^{abc}{\rm F}_{1}^{a} {\rm F}_{2}^{b} {\rm F}_{3}^{c} ,
\label{e:f}\cr }
\end{eqnarray}
where
${\rm F}^{a} = {1 \over 2} \lambda^{a}$ is the quark colour charge, 
the lower index 
indicates the number of the quark, $\lambda^{a}$ are the Gell-Mann 
matrices, and $f^{abc}$, $d^{abc}$ are the SU(3) structure constants.
Only the first two factors, Eqs. (\ref{e:sum},\ref{e:d}) are SU(3) 
invariants, however, i.e., only 
they can be expressed in terms of Casimir operators as follows
\begin{eqnarray}
\sum_{i < j}^{3} {\rm F}_{i} \cdot {\rm F}_{j} &=& {1 \over 2} 
C_{i + j + k}^{(1)} - 2 \\
\label{e:sum1}
d^{abc}{\rm F}_{1}^{a} {\rm F}_{2}^{b} {\rm F}_{3}^{c} &=& 
{1 \over 6} \left[C_{i + j + k}^{(2)} - 
{5 \over 2}C_{i + j + k}^{(1)} + {20 \over 3}\right]; \
\label{e:d1}
\end{eqnarray}
where $i + j + k$ stands for the 
three-quark colour state. 
Only the second factor, Eq. (\ref{e:d1}), depends on the cubic 
Casimir operator.
The third colour factor, Eq. (\ref{e:f}), is an off-diagonal operator 
that annihilates the two SU(3) eigenstates with definite exchange 
symmetry, i.e. 
the {\bf 1} and {\bf 10}, see Table \ref{table:3qstates}, and 
converts one 
{\bf 8} state into another. Therefore, it is not allowed in the quark 
model Hamiltonian
\footnote{It is true, of course, that the ``Mercedes-Benz star" 
diagram of QCD carries this colour factor.
This factor does not appear in the quark model because that diagram 
alone is not gauge invariant.}.

The first two colour factors Eqs. (\ref{e:sum},\ref{e:d}) are 
symmetric under the interchange 
of any pair of indices $i \leftrightarrow j$, $i \leftrightarrow k$, 
and
$j \leftrightarrow k$, whereas the third one, Eq. (\ref{e:f}), is 
antisymmetric.
All three are symmetric under cyclic permutations $i \rightarrow j 
\rightarrow k$
and $i \rightarrow k \rightarrow j$. Since the complete potential has 
to be symmetric under each of these permutations, the corresponding 
spin-spatial parts have to have the same symmetry properties as the 
colour ones. Consequently, the third one has to be spin dependent, 
whereas the first two need not. 
For the sake of simplicity in this letter we limit ourselves to 
spin-independent potentials, i.e. again to the first two cases, Eqs. 
(\ref{e:sum},\ref{e:d}).

Keeping with the tradition of the quark model, we take 
the harmonic oscillator for both the two- and three-quark spatial 
parts of potentials:
\begin{eqnarray}
{\cal V}_{12} &=& {1 \over 2} m \omega^2 \left({\bf r}_{1} - {\bf 
r}_{2}\right)^2
\label{e:pot12} \\
{\cal V}_{123} &=& c {1 \over 2} m \omega^2 
\left[\left({\bf r}_{1} - {\bf r}_{2}\right)^2 + 
\left({\bf r}_{3} - {\bf r}_{2}\right)^2 + 
\left({\bf r}_{1} - {\bf r}_{3}\right)^2 \right];
\label{e:pot123} \
\end{eqnarray}
with an as yet undetermined strength $c$ for the latter.
With the harmonic oscillator assumption we find that the 
${\rm F}_{i} \cdot {\rm F}_{j}$ model two-body interaction leads to 
the same form
of the 
effective potential in the $q^3$ system as the three-body force Eq. 
(\ref{e:sum}).
Similar statements hold for the colour-independent two- and 
three-body potentials. 
For this reason there is no need to 
introduce such two- and three-body potentials separately, but only 
one of a kind. 
We shall show that
a colour-independent two-body potential is necessary for the 
stabilization of
both $q {\bar q}$ and $q^3$ spectra. Hence we do not introduce a 
separate
colour-independent three-body potential.
With these results we can write down the Hamiltonians for few-quark 
systems
and then solve for their spectra.

\subsection{$q^3$ Hamiltonian and its spectrum} 

We shall start with the
${\rm F}_{i} \cdot {\rm F}_{j}$ model two-body potential and show 
that the ``cubic Casimir" 3-body force alone cannot stabilize it.
We find the following Hamiltonians
in the colour channels of the $q^3$ system 
\begin{eqnarray}
H_{1} &=& \sum_{i}^{3} {{\bf p}_{i}^2 \over {2m}} + 
{2 \over 3} \sum_{i < j}^{3} {\cal V}_{ij} + 
{10 \over 9} {\cal V}_{123}
\label{e:ham1} \\
H_{8} &=& \sum_{i}^{3} {{\bf p}_{i}^2 \over {2m}} +
{1 \over 6} \sum_{i < j}^{3} {\cal V}_{ij} - 
{5 \over 36} {\cal V}_{123}
\label{e:ham8} \\
H_{10} &=& \sum_{i}^{3} {{\bf p}_{i}^2 \over {2m}} - 
{1 \over 3} \sum_{i < j}^{3} {\cal V}_{ij} + 
{1 \over 9} {\cal V}_{123}
\label{e:ham10} .\
\end{eqnarray}
After going to centre-of-mass and Jacobi coordinates 
$\mbox{\boldmath$\rho$},
\mbox{\boldmath$\lambda$}$
we find the following  
potentials 
\begin{eqnarray}
V_{1} &=& 
\left({2 \over 3} + {10 \over 9}c \right)
{3 \over 2} m \omega^2 \left(\mbox{\boldmath$\rho$}^2 + 
\mbox{\boldmath$\lambda$}^2 \right)
\label{e:v1} \\
V_{8} &=& 
\left({1 \over 6} - 
{5 \over 36} c \right)
{3 \over 2} m \omega^2 \left(\mbox{\boldmath$\rho$}^2 + 
\mbox{\boldmath$\lambda$}^2 \right)
\label{e:v8} \\
V_{10} &=& 
\left(- {1 \over 3}  + 
{1 \over 9} c \right)
{3 \over 2} m \omega^2 \left(\mbox{\boldmath$\rho$}^2 + 
\mbox{\boldmath$\lambda$}^2 \right)
\label{e:v10} \
\end{eqnarray}
from which we can read off the stability conditions as
\begin{eqnarray}
1 &>& - {5 \over 3}c
\label{e:ineq1} \\
1 &>& {5 \over 6}c
\label{e:ineq8} \\
1 &<& {1 \over 3}c
\label{e:ineq10}. \
\end{eqnarray}
Note that two of the three inequalities are in conflict, regardless 
of the sign of $c$.
We conclude that the cubic Casimir three-body force cannot stabilize 
the 3q system
with the ${\rm F}_{i} \cdot {\rm F}_{j}$ model two-quark interaction.
Consequently, we turn to modification of this model that will lead to 
stable states in both the $q \bar q$ and $q^3$ systems.

Sufficient condition for the stabilization of the {\bf 8} $q \bar q$ 
state is to have as the two-body potential
\begin{eqnarray}
V_{12} &=& \left[c_{1} + {4 \over 3} + {\rm F}_{i} \cdot {\rm 
F}_{j}\right]
{1 \over 2} m \omega^2 \left({\bf r}_{1} - {\bf r}_{2}\right)^2
\label{e:newpot12} ,\
\end{eqnarray}
with $c_1 > 0$. For simplicity we take $c_1 = 1$. With this two-body 
potential 
we find
\begin{eqnarray}
V_{1} &=& 
\left({5 \over 3} + {10 \over 9}c \right)
{3 \over 2} m \omega^2 \left(\mbox{\boldmath$\rho$}^2 + 
\mbox{\boldmath$\lambda$}^2 \right)
\label{e:nv1} \\
V_{8} &=& 
\left({13 \over 6} - 
{5 \over 36} c \right)
{3 \over 2} m \omega^2 \left(\mbox{\boldmath$\rho$}^2 + 
\mbox{\boldmath$\lambda$}^2 \right)
\label{e:nv8} \\
V_{10} &=& 
\left({8 \over 3}  + 
{1 \over 9} c \right)
{3 \over 2} m \omega^2 \left(\mbox{\boldmath$\rho$}^2 + 
\mbox{\boldmath$\lambda$}^2 \right)
\label{e:nv10} \
\end{eqnarray}
Note that with this new and improved two-body interaction and 
{\it no} three-body force ($c = 0$), all three $q^3$ colour states
are stable, but the octet {\bf 8} is lighter than the 
singlet {\bf 1}, again in contrast with the experiment!

Turning on the three-body force, $c \neq 0$, we find the following 
stability condition
\begin{eqnarray}
- {3 \over 2} < c < {78 \over 5}
\label{e:nineq1} . \
\end{eqnarray}
For values of $c < {2 \over 5}$ we find the anticipated ordering of 
colour states: singlet {\bf 1} is the lowest lying, the next lowest is 
the octet {\bf 8}, and then the decimet {\bf 10}. The ratio of their 
ground state
energies can be made arbitrarily large by choosing $c$ sufficiently
close to - 1.5. For example, with $c$ = - 1.43, the {\bf 8} and 
{\bf 10} states are lying above 4 GeV.
We conclude that the colour-independent two-body force stabilizes 
the $q^3$ system, whereas the cubic Casimir three-body force makes 
it well ordered in colour, i.e. properly confined.

This should not be a surprise: for the colour singlet 
to separate away from other colour multiplets, the Hamiltonian must
contain at least one piece proportional to the colosinglet projection 
operator $P_{1}$ shown below
\begin{mathletters}
\begin{eqnarray}
{\rm P}_{1} &=& {1 \over 27} - {1 \over 36}
\sum_{i < j}^{3} \mbox{\boldmath$\lambda$}_{i} \cdot 
\mbox{\boldmath$\lambda$}_{j}
 + {1 \over 12} d^{abc} \mbox{\boldmath$\lambda$}_{1}^{a} 
\mbox{\boldmath$\lambda$}_{2}^{b} \mbox{\boldmath$\lambda$}_{3}^{c} 
\label{e:1su3}\\
{\rm P}_{8} &=& {16 \over 27} - {1 \over 9}
\sum_{i < j}^{3} \mbox{\boldmath$\lambda$}_{i} \cdot 
\mbox{\boldmath$\lambda$}_{j}
 - {1 \over 6} d^{abc} \mbox{\boldmath$\lambda$}_{1}^{a} 
\mbox{\boldmath$\lambda$}_{2}^{b} \mbox{\boldmath$\lambda$}_{3}^{c} 
\label{e:8su3}\\
{\rm P}_{10} &=& {10 \over 27} + {5 \over 36}
\sum_{i < j}^{3} \mbox{\boldmath$\lambda$}_{i} \cdot 
\mbox{\boldmath$\lambda$}_{j}
 + {1 \over 12} d^{abc} \mbox{\boldmath$\lambda$}_{1}^{a} 
\mbox{\boldmath$\lambda$}_{2}^{b} \mbox{\boldmath$\lambda$}_{3}^{c} 
\label{e:10su3}\
\end{eqnarray}
\end{mathletters}
which manifestly depends on both the two-body and the three-body
operators. Without the three-quark (cubic Casimir), there is bound 
to be some (accidental) degeneracy left and the colour singlet state
could not be isolated. 

Concrete (observable) phenomenological consequences of our new three-quark 
interaction only become visible in (``exotic'') multiquark systems, such as 
$q^{4}{\bar q}$, or  $q^{2}{\bar q}^{2}$, because the ``ordinary'' states, 
such as $q^{3}$, 
only allow one colour singlet, and non-singlet states have not been 
observed. 

\section{The $q^2 {\bar q}^2$ system}
\label{model}

Having found a confining potential that predicts the presumed
ordering of the $q^{3}$ colour spectrum, we turn to its
application to the $q^2 {\bar q}^2$ system. 
We break up this system into three-body configurations.
For our purposes one can equivalently think of this system either 
as $(q^2 {\bar q})\bar q$, or as $q(q {\bar q}^2)$. Thus
we need to evaluate the three-body potential's colour factor 
in variously coloured $q^2 {\bar q}$ and $q {\bar q}^2$ states.

The $q^2 {\bar q}$ system can occupy one of the following four
colour states 
$\left({\bf 3} \otimes {\bf 3}\right) \otimes {\bf \bar{3}} = 
\left({\bf \bar{3}} \oplus {\bf 6}\right) \otimes {\bf \bar{3}} 
= \left({\bf \bar{6}} \oplus {\bf 3}_{a}\right) \oplus
\left({\bf 3}_{s} \oplus {\bf 15}\right)$.
Note that there are two colour triplets and that they have different 
symmetry properties 
under the interchange of the two quark indices: one is symmetric, 
another antisymmetric. This means that there will be two colour singlet
$q^2 {\bar q}^2$ states with corresponding quark interchange symmetry 
properties. Our three-body interaction will distinguish between the 
two colour singlet states. 

We must be careful about the definition of the colour factors
in the nonrelativistic three-body potential involving 
antiquarks as they are sensitive to the 
C-conjugation properties of the relativistic interaction from which 
the potential was derived (the latter's properties carry over
into the nonrelativistic limit for odd number of quarks). 
More specifically, one finds a difference between the Lorentz scalar 
and zeroth component of Lorentz vector models, which is unusual.
In the quark model one ordinarily replaces the
quark colour factor ${\rm F}^{a}$ by 
\begin{eqnarray}
{\bar {\rm F}}^{a} &=& - {1 \over 2} \mbox{\boldmath$\lambda$}^{aT}
= - {1 \over 2} \mbox{\boldmath$\lambda$}^{a*} \
\end{eqnarray}
which is the 
definition of the colour {\it charge} operator of an antiquark. 
Note, however, that the minus 
sign in this definition stems from the C-conjugation properties of the
{\it vector} current and not from SU(3) itself. 
So for Lorentz scalar, pseudoscalar and axial-vector 
interactions this sign changes into a plus. 
\footnote{This definition ignores
the SU(3) analog of G-parity transformation.} 
In two-body potentials this sign makes no difference, as there are two
such factors that cancel. In the three-body case the sign makes a difference. 
For example, due to the odd number 
of interacting particles there is a sign difference
between the ``cubic Casimir'' three-quark
and three-antiquark potentials in the Lorentz-vector model
(unlike the two-body potential), thus apparently violating 
C-conjugation symmetry and CPT, since P- and T- are conserved.  
For Lorentz scalar interactions the antiquark potential 
has the same sign as the quark potential and there is no such problem.
Lorentz-vector 3-point
functions are forbidden by C-conjugation (``Furry's theorem'') anyway, 
so we conclude that only Lorentz scalar three-quark potential is allowed.

Thus we conclude that the ``cubic Casimir'' three-body interaction 
must have the following colour factor when antiquarks are involved
\begin{eqnarray}
{\bar {\cal C}}_{123} { } = \cases{
- d^{abc}{\rm F}_{1}^{a} {\rm F}_{2}^{b} {\rm {\bar F}}_{3}^{c}  
\label{e:sum6} \cr
~~d^{abc}{\rm F}_{1}^{a} {\rm {\bar F}}_{2}^{b} {\rm {\bar F}}_{3}^{c}  
\label{e:d6}\cr
- d^{abc}{\rm {\bar F}}_{1}^{a} {\rm {\bar F}}_{2}^{b} {\rm {\bar F}}_{3}^{c}  
\label{e:d7}\cr}
\end{eqnarray}
Once again, we can express the two SU(3) invariant colour factors
in terms of the Casimir operators. The first one remains unchanged:
\begin{eqnarray}
\sum_{i < j}^{3} {\rm F}_{i} \cdot {\rm F}_{j} &=& 
{\rm F}_{1} \cdot {\rm F}_{2} + {\rm F}_{1} \cdot {\bar {\rm F}}_{3} +
{\rm F}_{2} \cdot {\bar {\rm F}}_{3}
= {1 \over 2} C_{i + j + k}^{(1)} - 2 ,
\label{e:sum3} \
\end{eqnarray}
whereas the second one becomes
\begin{eqnarray}
d^{abc}{\rm F}_{1}^{a} {\rm F}_{2}^{b} {\bar {\rm F}}_{3}^{c} &=& 
{1 \over 6} \left[C_{i + j + k}^{(2)} - 
{5 \over 2}C_{i + j}^{(1)} + {50 \over 9}\right] .
\label{e:d3} \
\end{eqnarray}

Note that in the second factor, Eq. (\ref{e:d3}) the first (quadratic) 
Casimir is 
evaluated between the two-quark (sub-)state $i + j$, which leads to a 
distinction between the two overall colour triplets 
(which are symmetric and antisymmetric in the quark indices).
This leads to results shown in Table \ref{table:2qq-barstates}.
Using Table \ref{table:2qq-barstates}, we find the following
potentials in the two overall colour singlet states [$s \equiv 8, a 
\equiv 1$]
\begin{eqnarray}
V_{s} &=& 
c {5 \over 18} \omega^2 
\left({\bf r}_{12}^2 + {\bf r}_{13}^2 + {\bf r}_{14}^2 +
{\bf r}_{23}^2 + {\bf r}_{24}^2 + {\bf r}_{34}^2 \right)
\label{e:3s} \\
V_{a} &=& 
- c {5 \over 9} \omega^2 
\left({\bf r}_{12}^2 + {\bf r}_{13}^2 + {\bf r}_{14}^2 +
{\bf r}_{23}^2 + {\bf r}_{24}^2 + {\bf r}_{34}^2 \right)
\label{e:3a}  .\
\end{eqnarray}
From the signs of the two interaction potentials we see that the 
mass/energy of the ``octet'' state is enhanced for $c \geq 0$ and vice 
versa for the ``singlet'' state. As we have already shown that $c$ 
can be either positive or negative (it only needs to be less than 0.4), 
we conclude that the 3-body interaction 
can elevate the mass of the unobserved (symmetric) ``octet'' states 
above the conventional/ordinary two-meson states
and thus make them less stable and less likely to be detected. 
In this sense, one may think of this three-body interaction as a
solution to Isgur's problem (``fiasco'') of unobserved (towers of) 
$q^2 {\bar q}^2$ states in the ${\rm F}_{i} \cdot {\rm F}_{j}$ model.

\section{Conclusions}

We have looked into the question of the second (cubic) Casimir
interaction in the quark model and found that:

\begin{enumerate}
\item It is insufficient to stabilize the ${\rm F}_{i} \cdot {\rm 
F}_{j}$ model by itself. A colour-independent two-body force is 
necessary to prevent colour dissolution
in both the $q \bar q$ and the $q^3$ systems.
\item In conjunction with a colour-independent two-body force
it leads to a proper ordering in energy of the coloured $q^3$ states
for three-body coupling constants $c$ in the range $- 1.5 < c < 0.4$.
\item In the $q^2 {\bar q}^2$ system 
it leads to a distinction between the two colour singlet states, 
in that it enhances overall binding in one and diminishes it in the 
other, depending on the sign of its coupling constant $c$.
\end{enumerate}

We have made several simplifying assumptions that can and ought to
be relaxed in the future. For example: (1) harmonic oscillator nature-, 
and (2) spin independence of the three-body potential.
Relaxation of these assumptions leads to new  
predictions in the observable (colour singlet) sector.
For example, the replacement of the usual two-body ``colour-spin'' 
interaction with a more complicated, three-body one will change 
the pattern of SU(6) splitting in multi-quark states. 

Many papers have been written on the ``saturation'' of 
quark-quark interactions, starting with those of Nambu 
\cite{Nambu} and of Greenberg and Zwanziger \cite{Greenberg}.
Those early papers assumed two-quark (quadratic Casimir operator) 
potentials that vanish at infinite 
quark-quark separations, in contrast with modern notions that allow 
them to infinitely rise.
This work can be viewed as a natural continuation of those early works 
to models with infinitely 
rising potentials and three-quark (cubic Casimir) colour operators.

\newpage

\begin{table}
\caption{Diagonal matrix elements of the three operators for 
variously coloured $q^3$ states. Of course, there are two distinct 
{\bf 8} states, but they are equivalent in this regard.
\label{table:3qstates}}
\begin{tabular}{lllll} & {~~~{\bf 1}} & {~~~{\bf 8}} & {~~{\bf 10}} \\
 \tableline
 &&& \\
{$\langle \sum_{i < j}^{3} {\rm F}_{i} \cdot {\rm F}_{j}  \rangle$} & 
{~$- 2$} & { - $\frac{1}2$} & {~~~1}  \\
&&& \\
{$\langle d^{abc}{\rm F}_{1}^{a} {\rm F}_{2}^{b} {\rm F}_{3}^{c} 
\rangle$} & 
{~~$\frac{10}9$} & { - $\frac{5}{36}$} & {~~~$\frac{1}9$}  \\
&&& \\
{$\langle f^{abc}{\rm F}_{1}^{a} {\rm F}_{2}^{b} {\rm F}_{3}^{c} 
\rangle$} & {~~~$0$} & {~~~$0$} & {~~~$0$} \\
&&& \\
\end{tabular}
\end{table}

\begin{table}
\caption{Diagonal matrix elements of the three-body colour operators 
for variously coloured $q^2 {\bar q}$ states.
\label{table:2qq-barstates}}
\begin{tabular}{llllll} 
& ${\bf 3}_{a}$ & ${\bf 3}_{s}$ & $\bar {\bf 6}$ &
{\bf 15}  
\\
\tableline
 &&&& \\
{$\langle \sum_{i < j}^{3} {\rm F}_{i} \cdot {\rm F}_{j}  \rangle$} 
 & -$\frac{4}{3}$ & -$\frac{4}{3}$ & - $\frac{1}3$ & $\frac{2}{3}$ \\
&&&& \\
{$\langle d^{abc}{\rm F}_{1}^{a} {\rm F}_{2}^{b} {\bar {\rm F}}_{3}^{c}  
\rangle$} 
& $\frac{5}9$ & - $\frac{5}{18}$ & $- \frac{5}{18}$ & $\frac{1}{18}$  
\\
&&&& \\
\end{tabular}
\end{table}

\end{document}